\documentclass[12pt,letterpaper]{JHEP3}
\usepackage{graphicx}
\usepackage{feynmf}
\usepackage{amsmath,amsfonts}

\title{Neutrino Masses, Leptogenesis, and Unification in the Absence of Low Energy Supersymmetry}

\author{Willy Fischler$^1$ and Raphael Flauger$^{1,2}$ \\
$^1${\it Department of Physics\\
     \hphantom{$^1$}University of Texas, Austin, TX 78712}\\

$^2${\it Kavli Institute for Theoretical Physics\\
  \hphantom{$^2$}University of California, Santa Barbara, CA 93106}\\

E-mail: \email{fischler@physics.utexas.edu}\\
\hphantom{E-mail:}        \email{flauger@physics.utexas.edu}}

\abstract{We explore generating the dimension five operator leading to neutrino masses by integrating out heavy $SU(2)_L $ triplet fermions. We exhibit a model that has a neutrino mass matrix consistent with observations. In addition, the model is capable of producing the observed baryon asymmetry, and unification can be achieved without dangerous proton decay. Supersymmetry can be broken at the fundamental scale with no supersymmetric signal at low energies.}


\preprint{UTTG-05-08 \\NSF-KITP-08-70}

\begin{document}


\section{Introduction}

A remarkable collection of observations of neutrino oscillations confirms, if there was any need to, the existence of physics beyond the standard model. What precisely the new degrees of freedom are that may lurk beyond the standard model, and what their interactions are, is still a mystery. We all have our prejudices about what lies beyond the standard model, but in the final analysis only observations will inform us of physics beyond the energies we have so far been limited to.  
The observations leading to the neutrino mass matrix do give us some clues, but at the same time they leave us with tremendous freedom to speculate about the precise microscopic physics responsible for neutrino masses.  \\

The gauge invariant dimension five operator that is thought to give rise to neutrino masses once the electroweak symmetry is broken could be generated at tree level or at loop level.
Assuming it is generated at tree level, we know that it has three different realizations~\cite{Ma:1998dn}. The three cases can be thought of as extensions of the standard model by a very massive gauge-singlet fermion, a very massive scalar transforming in the adjoint of $SU(2)_L$ carrying hypercharge, and a very massive fermion transforming in the adjoint of $SU(2)_L$ carrying no hypercharge. All these degrees of freedom might be present in nature, with masses beyond the reach of our current experiments. In this case each of them would contribute to the neutrino masses. It seems reasonable, however, to assume that the contribution of one of them will dominate over the others. The cases where the dominant contribution is due to the heavy singlet fermion and the heavy scalar triplet have been studied extensively in the literature. \\

In this paper, we will explore some of the possible implications that arise in the case that the dominant contribution comes from triplet fermions~\cite{Foot:1988aq}, \cite{Ma:1998dn}, \cite{Ma:2002pf}.

The decay of these triplets violates lepton number. In models with at least three generations of triplet fermions it is generically CP violating, and it is slow enough to lead to a small departure from thermal equilibrium. In other words, the prerequisites for a successful generation of the baryon asymmetry are met. Detailed calculations show that a model with three families of triplet fermions is indeed capable of producing the observed baryon asymmetry for triplet masses above some $10^{10}\,GeV$.\footnote{During the final stages of this work, we became aware through reference~\cite{Franceschini:2008pz} that the earlier paper \cite{Hambye:2003rt} has some overlap with our Section \ref{lepto}.}

The addition of these triplet fermions modifies the running of the standard model gauge couplings. We will be prejudiced and demand that the gauge couplings of the standard model unify. Consistent with this bias, we will include grand unification partners to the triplet fermions. Clearly, the precise field content will depend on a choice of GUT group. To give an explicit realization of our ideas we will consider embedding the standard model gauge group into $SU(5)$ but the idea works just as well for the other popular GUT groups. \\

At low energies, the model we will exhibit in this paper therefore just has the standard model degrees of freedom, with neutrino masses and mixings encoded in a dimension five operator. 
At intermediate scales, the microscopic physics we will invoke as being responsible for this dimension five operator will require three families of triplet fermions. When put into the context of grand unified theories there will also be octet fermions at intermediate scales and additional degrees of freedom at the GUT scale, but these  heavy degrees of freedom will not play a role for most aspects we will discuss. \\

Since the model generically only contains the standard model degrees of freedom at low energies, it cannot possibly address the gauge hierarchy problem. In this regard we take the point of view that our so far trustworthy guide, naturalness, may well let us down when it comes to estimating the magnitude of the cosmological constant or even the Higgs mass.\footnote{For a detailed and thoughtful discussion see~\cite{ArkaniHamed:2004fb}} There may well be some explanation for the smallness of these parameters in our low energy effective field theories that has escaped us so far. The other option, unsatisfying as it may seem, is that the explanation for the smallness of the cosmological constant and (or) the Higgs mass is anthropic. \\

The organization of this paper is as follows: In the next section, we will exhibit the model that will be starting point for our calculations in Sections \ref{lmass} and \ref{lepto}.

In Section \ref{lmass}, we present the mass matrix for the light neutrinos and mention bounds on the masses of the triplet fermions imposed by experimental limits on rare decays, experimental limits on the electric dipole moment of the electron, and electroweak precision measurements. These bounds turn out to be rather weak.

By solving the Boltzmann equations numerically, we show in Section \ref{lepto} that, for triplet masses around $10^{10}\,GeV$, the model can generate the observed baryon asymmetry. A rather nice feature of this model is that the triplets' gauge interactions ensure that the initial conditions for the Boltzmann equations are given by thermal equilibrium, something that is not at all clear in the case of leptogenesis via the more commonly studied singlet fermions.

In Section \ref{unification} we present an explicit realization in the context of an $SU(5)$ GUT theory. We derive bounds on the masses of triplets and their unification partners imposed by our bias that unification should occur. We find that, assuming all three triplet masses are of the same order of magnitude, unification requires the triplets to have masses around $10^{10}\,GeV$, which incidentally coincides with what was required for successful generation of the baryon asymmetry. 

In the conclusions, we mention an extension of our ideas that gives rise to a candidate for dark matter and a different way of obtaining the baryon asymmetry of the universe. 
\newpage
\section{The Model and the Actors I}

In addition to the standard model degrees of freedom, the model has three families of Weyl fermions, 
$T_i$, where $i=1,2,3$ denote the families, that transform as singlets under $SU(3)_c$, as triplets under $SU(2)_L$, and carry no hypercharge. We will refer to these as ``triplets''.\footnote{We can, of course, equivalently think of them as Majorana fermions.}
With the standard model extended by these new degrees of freedom, the most general renormalizable Lagrangian takes the form
\begin{multline}
\mathcal{L}=\mathcal{L}_{\text{SM}}-T^{\alpha A}_i i\partial_{\alpha \dot\alpha}\overline{T}^{\dot\alpha A}_i-\frac12M_i\left(T^{\alpha A}_iT_{i\alpha}^A+\overline{T}_{i\dot\alpha}^{ A}\overline{T}_i^{\dot\alpha A}\right)\\-y_{ij}\epsilon_{ac}{{\tau^A}^a}_b\ell_i^{\alpha c}T_{j\alpha }^A H^{\dagger b}+y^*_{ij}\epsilon^{ac}{{\tau^A}^b}_a\overline\ell_{i\dot\alpha c}\overline{T}_{j}^{\dot\alpha A} H_{ b} \,.
\end{multline}
Some remarks on our conventions are in order. The fields $T$, $\ell$, and $H$ are the triplet, lepton doublet, and standard model Higgs doublet, respectively. Lower case indices from the beginning of the Roman alphabet are $SU(2)_L$ indices with superscripts denoting the fundamental representation and subscripts denoting the anti-fundamental representation. Lower case letters from the middle of the Roman alphabet label the three different families. Upper case letters label objects transforming in the adjoint representation of $SU(2)_L$. Lower case letters from the beginning of the Greek alphabet are the usual $SL(2,\mathbb{C})$ spinor indices. At this point, the triplet masses, $M_i$ are still arbitrary as are the Yukawa couplings, $y_{ij}$. We will see how those are related to the masses of the light neutrinos in Sections \ref{numass}. We take the $SU(2)_L$ generators, ${{\tau^A}^a}_b$, to be normalized according to $Tr(\tau^A\tau^B)=\frac12\delta^{AB}$. 

\section{Masses of the Standard Model Leptons}\label{lmass}
As was first discussed in \cite{Foot:1988aq}, the presence of triplet states leads to neutrino masses. The resulting neutrino mass matrix takes the same form as in the case of singlets, which means that the two models are equally viable.\\ 
In the charged lepton sector the two models differ significantly. The presence of triplet states potentially leads to flavor changing neutral currents and an electric dipole moment for the electron, but the smallness of the neutrino masses guarantees that the model is far from being in conflict with current experimental bounds. \\
Because of the vector-like nature of the triplets, electroweak precision data does not lead to any interesting constraints on the model. 

\subsection{Neutrino Masses}\label{numass}
Let us first turn our attention to the generation of the light neutrino masses. At tree-level, the only operators that are needed for our discussion are
\begin{equation}
\mathcal{L}\supset-\frac12M_i\left(T^{\alpha A}_iT_{i\alpha}^A+ \overline{T}_{i\dot\alpha}^{ A}\overline{T}_i^{\dot\alpha A}\right) -y_{ij}\epsilon_{ac}{{\tau^A}^a}_b\ell_i^{\alpha c}T_{j\alpha }^A H^{\dagger b}+y^*_{ij}\epsilon^{ac}{{\tau^A}^b}_a\overline\ell_{i\dot\alpha c}\overline{T}_{j}^{\dot\alpha A} H_{ b} \,.
\end{equation}
Once the Higgs field acquires a vacuum expectation value, $\langle H_a\rangle=\frac{1}{\sqrt{2}}v\delta_{a}^2$,\footnote{To be specific, in these conventions we have $v=246GeV$.} this leads to a mass matrix for the neutral components of leptons and triplets of the form
\begin{equation}
\mathcal{L}\supset-\frac12\left(\begin{array}{cc}\ell_i^{1\alpha} &T_i^{\alpha3}\end{array}\right)\left(\begin{array}{cc}0&-\frac{1}{2\sqrt{2}}{y}_{ij}v\\-\frac{1}{2\sqrt{2}}{y}^T_{ij}v&M_i\delta_{ij}\end{array} \right)\left(\begin{array}{c}\ell_j^{1\alpha}\\ T_j^{\alpha 3}\end{array} \right)\, + h.c.
\end{equation}
Up to an overall factor of $1/2$, which we might as well absorb into the Yukawa couplings, the mass matrix, 
\begin{equation}
\mathcal{M}=\left(\begin{array}{cc}0&-\frac{1}{2\sqrt{2}}{y}_{ij}v\\- \frac{1}{2\sqrt{2}}{y}^T_{ij}v&M_i\delta_{ij}\end{array} \right)\,,
\end{equation}
is the same as in the more commonly studied case of heavy right-handed gauge singlet fermions. It is therefore apparent that the triplet model is equally viable as the singlet model, at least as far as the mass matrix for the light neutrinos is concerned. The diagonalization and parametrization of the mass matrix of the light neutrinos in terms of mixing angles proceeds in the same way as in the singlet model and the reader familiar with the discussion in the singlet case can safely skip the rest of this subsection. We merely include it to establish the notation we use in the rest of the paper. \\
The mass matrix is complex symmetric, and we know from the singular value decomposition of complex symmetric matrices that it is related to a diagonal matrix with real entries on the diagonal according to
\begin{equation}
\mathcal{M}=\mathcal{U^*}\left(\begin{array}{cc}D_m& 0\\
0&D_M\end{array} \right)\mathcal{U}^\dagger\,,
\end{equation}
where $\mathcal{U}$ is a $6\times 6$ unitary matrix, $D_m=\text{diag}(m_1,m_2,m_3)$ are the masses of the light neutrinos, and $D_M=\text{diag}(\tilde{M}_1,\tilde{M}_2,\tilde{M}_3)$ are the masses of the neutral components of the triplets. Introducing $m_{D\,ij}\equiv -\frac{1}{2\sqrt{2}}y_{ij}v$, and assuming that all the entries in $m_D$ are much smaller than the masses of the triplets, the diagonalization can be done in perturbation theory. Introducing fields $\nu_{i\alpha}$, the flavor eigenstates of the light neutrinos, and $\tilde{T}_{i\alpha}$, the neutral components of the triplets, that are related to $\ell_i^{1\alpha}$ and $T_i^{\alpha3}$ to leading order in $m_D/M$ by
\begin{equation}
\left(\begin{array}{c}\ell^{1\alpha}\\ T^{\alpha 3}\end{array} \right)=\left(\begin{array}{cc}1& m_D^*M^{-1}\\-M^{-1}m_D^T&1\end{array}\right)\left(\begin{array}{c}\nu^{\alpha}\\ \tilde{T}^{\alpha}\end{array} \right)\,,
\end{equation}
the mass terms of neutrinos and triplets take the form\footnote{$M_i$ and $\tilde{M}_i$ are equivalent to this order. }
\begin{equation}
\mathcal{L}\supset -\frac12 \nu_i^\alpha m_{\nu\,ij}\nu_{j\alpha}-\frac12 M_i\tilde{T}_i^\alpha\tilde{T}_{i\alpha}+h.c.\,,
\end{equation}
with the mass matrix for the light neutrinos, $m_\nu$, given by the familiar formula
\begin{equation}
m_\nu=-m_DM^{-1}m_D^T\,.
\end{equation}
It can be diagonalized as usual by a $3\times 3$ unitary matrix $U_0$ relating flavor and mass eigenstates according to
\begin{equation}\label{eq:diagnumass}
D_m=-U_0^T m_D M^{-1} m_D^T U_0\,.
\end{equation}
For completeness, let us note that this implies that the unitary matrix $\mathcal{U}$ is given by
\begin{equation}
\mathcal{U}=\left(\begin{array}{cc}1& m_D^*M^{-1}\\-M^{-1}m_D^T&1\end{array}\right)\left(\begin{array}{cc}U_0& 0\\0&1\end{array}\right)\,.
\end{equation}
It will be convenient for later sections to express the Yukawa couplings $y_{ij}$ in terms of the masses of light neutrinos and neutral components of the triplets. From equation~\eqref{eq:diagnumass} we see that the matrix
\begin{equation}
R\equiv i\sqrt{D_M}^{-1}m_D^TU_0\sqrt{D_m}^{-1}
\end{equation} 
is complex orthogonal, {\it i.e.} satisfies $R^TR=1$. In terms of this matrix the Yukawa couplings are related to the masses of neutrinos and triplets according to:
\begin{equation}
y=\frac{2\sqrt{2}i}{v}U_0^*\sqrt{D_m}R^T\sqrt{D_M}\,.
\end{equation}

\subsection{Charged Lepton Masses and Experimental Constraints on the Model}
Let us now turn to the masses of the charged leptons. Here we encounter the first qualitative difference between the singlets and the triplets because the charged components of the triplets can mix with the charged leptons. The relevant part of the Lagrangian for this part of our discussion is 
\begin{equation}
\mathcal{L}\supset-\frac12M_iT^{\alpha A}_iT_{i\alpha}^A -y_{ij}\epsilon_{ac}{{\tau^A}^a}_b\ell_i^{\alpha c}T_{j\alpha }^A H^{\dagger b}-y_{e,ij}\ell_i^{\alpha a} H_a \tilde{e}_{j\alpha}^c+h.c. \,,
\end{equation}
where the superscript $c$ in $\tilde{e}_{j\alpha}^c$ denotes charge conjugation. 
In the broken phase this leads to a mass term in the Lagrangian for the charged leptons of the form
\begin{equation}
\mathcal{L}\supset-\left(\begin{array}{cc}\ell_i^{2\alpha} &T_i^{\alpha-}\end{array}\right)\left(\begin{array}{cc}\frac{{y}_{e,ij}v}{\sqrt{2}}&-\frac{{y}_{ij}v}{2}\\0&M_i\delta_{ij}\end{array} \right)\left(\begin{array}{c}\tilde{e}_{j\alpha}^{c}\\ T_{j\alpha}^{+}\end{array} \right)+h.c.
\end{equation}
We can perform a singular value decomposition to obtain the mass eigenvalues and eigenstates using unitary matrices $\mathcal{V}$ and $\mathcal{W}$ according to
\begin{equation}
\mathcal{M}=\left(\begin{array}{cc}\frac{{y}_{e,ij}v}{\sqrt{2}}&-\frac{{y}_{ij}v}{2}\\0&M_i\delta_{ij}\end{array} \right)=\mathcal{V}^*\left(\begin{array}{cc}\tilde{m}_{i}\delta_{ij}&0\\0&\hat{M}_i\delta_{ij}\end{array} \right)\mathcal{W}^\dagger\,,
\end{equation}
where $\tilde{m}_{i}$ are $m_e$, $m_\mu$, and $m_\tau$, the masses of the charged leptons and $\hat{M}_i$ are the masses of the charged triplets. The unitary matrices $\mathcal{V}$ and $\mathcal{W}$ can again easily be obtained perturbatively. To second order they are
\begin{equation}
\mathcal{V}=\left(\begin{array}{cc}1-m_D^*M^{-2}m_D^T&\sqrt{2} m_D^*M^{-1}\\-\sqrt{2}M^{-1}m_D^T&1-M^{-1}m_D^Tm_D^*M^{-1}\end{array}\right)\left(\begin{array}{cc}V_l& 0\\0&V_T\end{array}\right)\,,
\end{equation}
and 
\begin{equation}
\mathcal{W}=\left(\begin{array}{cc}1&\sqrt{2} m_e^\dagger m_DM^{-2}\\-\sqrt{2}M^{-2}m_D^\dagger m_e&1\end{array}\right)\left(\begin{array}{cc}W_l& 0\\0&W_T\end{array}\right)\,,
\end{equation}
where we have introduced $m_e={y}_{e}v/\sqrt{2}$, and $V_l$, $W_l$ are the unitary $3\times3$ matrices that are usually used to diagonalize the lepton Yukawa couplings.\\
This mixing will in general lead to electric dipole moments for the electron as well as to rare decays such as $\mu^-\to e^-e^-e^+$ at tree level, and $\mu^-\to e^-\gamma$ at one loop. This can be seen by looking at the coupling of the light leptons and triplets to the gauge bosons. The photon couples universally to leptons and charged triplet components, so its couplings will still be diagonal. There will be mixing in the couplings of leptons to the W-boson. Some of these are already present in the form of the PMNS matrix if neutrino masses are due to heavy singlets, but there are also couplings to the W-boson mixing light leptons and triplets, which are not present in the singlet case. These potentially lead to electric dipole moments of the electron. These types of couplings between leptons and triplets also arise in couplings to the Z-boson, where in addition we also have potentially dangerous mixings between the light leptons giving rise to flavor changing neutral currents. Since the couplings to the W-boson do not give rise to any dangerous effects that are not already present in the couplings to the Z, it is then sufficient, at least for the purposes of constraining the triplet masses, to focus on the couplings of charged leptons and charged triplets to the Z-boson. 
For this discussion the relevant term in the Lagrangian is
\begin{equation}
\mathcal{L}\supset-\ell^{\alpha a}_i iD_{\alpha \dot\alpha}\overline{\ell}^{\dot\alpha}_{a\,i}-\tilde{e}^{c\,\alpha}_i iD_{\alpha \dot\alpha}\overline{\tilde{e}}^{c\,\dot\alpha }_i -T^{\alpha A}_i iD_{\alpha \dot\alpha}\overline{T}^{\dot\alpha A}_i
\end{equation}
If we introduce the fields corresponding to mass eigenstates of the charged leptons and charged triplets according to
\begin{equation}
\left(\begin{array}{c}\ell^{2\alpha}\\ T^{\alpha -}\end{array} \right)=\left(\begin{array}{cc}1-m_D^*M^{-2}m_D^T&\sqrt{2} m_D^*M^{-1}\\-\sqrt{2}M^{-1}m_D^T&1-M^{-1}m_D^Tm_D^*M^{-1}\end{array}\right)\left(\begin{array}{cc}V_l& 0\\0&V_T\end{array}\right)\left(\begin{array}{c}e^{\alpha}\\ \tilde{T}^{\alpha -}\end{array} \right)\,,
\end{equation}
and similarly
\begin{equation}
\left(\begin{array}{c}\tilde{e}^{c}_{\alpha}\\ T^{+}_{\alpha}\end{array} \right)=\left(\begin{array}{cc}1&\sqrt{2} m_e^\dagger m_DM^{-2}\\-\sqrt{2}M^{-2}m_D^\dagger m_e&1\end{array}\right)\left(\begin{array}{cc}W_l&0\\0&W_T\end{array}\right)\left(\begin{array}{c}e^{c}_{\alpha}\\ \tilde{T}^{+}_{\alpha}\end{array} \right)\,,
\end{equation}
the coupling of the charged leptons and charged triplets to the Z-boson up to order $\mathcal{O}(m^2_{e,D}/M^2)$ takes the following form
\begin{eqnarray}
\mathcal{L}&&\supset
\left(\begin{array}{cc}\overline{e}^c_{i\,\dot\alpha} &\tilde{\overline{T}}_{i\dot\alpha}^{+}\end{array}\right)\left(\begin{array}{cc}\frac{{g'}^2}{\sqrt{g^2+{g'}^2}}&\frac{2\sqrt{g^2+{g'}^2}W_l^\dagger m_e^\dagger m_D^*M^{-2}W_T}{\sqrt{2}}\\\frac{2\sqrt{g^2+{g'}^2}W_T^\dagger M^{-2}m_D^\dagger m_e W_l}{\sqrt{2}}&-\frac{g^2}{\sqrt{g^2+{g'}^2}}\end{array}\right)
\left(\begin{array}{c}e_{j\,\alpha}^{c}\\ \tilde{T}_{j\,\alpha}^{+}\end{array}\right)Z^{\dot\alpha \alpha}+\nonumber\\\nonumber
&&\left(\begin{array}{cc}\overline{e}_{i\,\dot\alpha} &\tilde{\overline{T}}_{i\dot\alpha}^{-}\end{array}\right)
\left(\begin{array}{cc}\frac{g^2-{g'}^2+2(g^2+{g'}^2)V_l^\dagger m_D^*M^{-2}m_D^TV_l}{2\sqrt{g^2+{g'}^2}}&-\frac{\sqrt{g^2+{g'}^2}V_l^\dagger m_D^*M^{-                1}V_T}{\sqrt{2}}\\-\frac{\sqrt{g^2+{g'}^2}V_T^\dagger M^{-1}m_D^TV_l}{\sqrt{2}}&\frac{g^2-{g'}^2V_T^\dagger M^{-1}m_D^Tm_D^*M^{-1}V_T}{\sqrt{g^2+{g'}^2}}\end{array}   \right)
\left(\begin{array}{c}e_{j\,\alpha}\\ \tilde{T}_{j\,\alpha}^{-}\end{array} \right)Z^{\dot\alpha \alpha}\\
\end{eqnarray}
Since the unitary matrices $V_{l,T}$ and $W_{l,T}$ are already determined by requiring that the mass matrix for the charged leptons and triplets be diagonal, we immediately see that the second contribution generically gives rise to flavor-changing neutral currents in the lepton sector leading for instance to $\mu^-\to e^-e^-e^+$ at tree level and $\mu^-\to e^-\gamma$ at one loop. We also see that there are generically non-vanishing CP-violating phases that give rise to an electric dipole moment for the electron. 
Both rare decays such as $\mu^-\to e^-e^-e^+$, or $\mu^-\to e^-\gamma$, and the electron electric dipole moment are of course strongly constrained experimentally. For any value of triplet masses that has not already been ruled out by the fact that no triplets were seen at LEP, the smallness of the light neutrino masses implies, however, that both rare decays and electric dipole moments generated in this model are beyond the reach of current experiments. This has been studied in more detail in~\cite{Abada:2007ux},~\cite{Abada:2008ea}.
Another stringent constraint on physics beyond the standard model comes from precision electroweak measurements. The triplets are charged under the weak force and it is natural to ask whether the masses of triplets can be bounded by precision electroweak data from LEP. Due to their vector-like nature this does not, however, yield a stronger constraint than the non-observation at LEP either. At leading order, the parameters $S$, $T$, $U$ as defined in the original paper~\cite{Peskin:1991sw} vanish identically. One might worry about contributions to $T$ and $U$ at higher loop order, but these are too small to constrain the triplets masses as well. Including into the study the parameters $X$, $V$, $W$ introduced in \cite{Maksymyk:1993zm} does not yield any interesting bounds on triplet masses, either. Instead, putting everything together, one finds that for light triplets the Higgs is allowed to be marginally heavier than usual.
We conclude that triplets of any mass are allowed that is not already ruled out by non-observation of triplets at LEP or the Tevatron. In particular, it is conceivable that triplets have just the right masses to be seen at the LHC. This requires tiny Yukawa couplings of the triplets to leptons and Higgs to keep the neutrino masses small. This implies that the decays of these light triplets then give rise to interesting signatures in the detector such as displaced vertices. This possibility has been studied very recently in more detail in \cite{Franceschini:2008pz}.\\
Since Yukawa couplings are protected by chiral symmetries, Yukawa couplings of any size, even though not necessarily esthetically pleasing, are technically natural.  Let us call the chiral symmetry protecting the Yukawa coupling between the triplets, the leptons, and the Higgs a $\mathbb{Z}_2$ symmetry for definiteness, but it could equally well be some other discrete group or even a continuous one. If the Yukawa couplings are very small, as is the case for light triplets, we have an approximate symmetry, and it seems natural to ask if one could turn this approximate symmetry into an exact one. In this case the triplets become a good dark matter candidate very much like the LSP. In the absence of the Yukawa coupling between the triplets and the Higgs and leptons, the neutrinos are exactly massless and it seems we have defeated the purpose of our model. One can, however, generate the neutrino masses at loop level if one introduces a scalar field with the same charge assignments under the standard model gauge group as the Higgs that is also charged under this symmetry. 
This may seem {\it ad hoc}, but especially in the context of grand unification this model has several nice properties such as an interesting mechanism for baryogenesis and a natural candidate for dark matter. We will leave this for a future publication.

\section{Generating the Baryon-Asymmetry of the Universe}\label{lepto}

A very nice feature of the most widely discussed seesaw mechanism involving right-handed gauge singlet fermions is that, in addition to giving neutrino masses, it also allows for the baryon asymmetry to be generated via leptogenesis \cite{Fukugita:1986hr}, \cite{Luty:1992un} (for a nice review and additional references see {\it e.g.}~\cite{Buchmuller:2004nz}). It is therefore natural to ask whether the analogue mechanism will still work in a seesaw mechanism that involves triplets.
We know that in order to generate the baryon asymmetry of the universe Sakharov's conditions have to be satisfied \cite{Sakharov:1967dj}. One of them is a departure from equilibrium while the asymmetry is generated. One might guess that because of the triplets' gauge interactions there will not be a sufficient departure from equilibrium and no asymmetry will be generated.  
However, despite the triplets' coupling to the $SU(2)_L$ gauge bosons, the observed baryon asymmetry can still be generated very much like in the original leptogenesis scenario \cite{Fukugita:1986hr} for what is believed to be the relevant range of neutrino masses. Our calculations show that the triplets give rise to the same amount of baryon asymmetry as the singlets. A nice feature of leptogenesis in models of neutrino masses involving triplets is that the calculations are independent of initial conditions, or rather the initial conditions are set by thermal equilibrium, because the gauge interactions bring the triplets close to thermal equilibrium rather quickly. We will show this in the remainder of this section, but before going into the details, let us briefly try to give an intuitive argument why all this is plausible. \\
As the temperature drops below the mass of the triplets their equilibrium abundance is Boltzmann suppressed. There are two main processes that drive the triplet abundance towards its equilibrium abundance: annihilation into $W$-bosons, and the decay of triplets into leptons and Higgs. While the decay in general is CP violating and generates a lepton asymmetry, no asymmetry is generated in the annihilation, which means there is a competition between the two channels. The coupling controlling the annihilation is the gauge coupling, which is typically bigger than the Yukawa coupling controlling the decay. However, for temperatures below the triplet mass the annihilation rate per triplet is Boltzmann suppressed since it is proportional to the number of triplets. This makes up for the smallness of the Yukawa couplings. For small neutrino masses the Yukawa couplings controlling the decays and inverse decays are small, and the annihilation is the dominant process. In this regime the triplets are less efficient than singlets (assuming that the singlets were in thermal equilibrium, which is not necessarily a good assumption for small Yukawa couplings). For larger values of neutrino masses the Yukawa couplings are larger and the decay dominates, leading to an asymmetry, until they get so large that decays and inverse decays are in thermal equilibrium and any asymmetry gets washed out. In this regime the triplets behave just like singlets except for a factor of three because in thermal equilibrium there are three triplets per singlet. This factor of three is canceled by a factor of one third appearing in the CP violation for the triplets as compared to the CP violation for singlets, and in the end we find that singlets and triplets generate the same baryon asymmetry for the relevant range of neutrino masses.  To determine what values of neutrino masses are large or small according to this criterion requires detailed study, and we will see that for the range of neutrino masses that seems to be realized in nature, just like for the singlets, leptogenesis is dominated by decays and inverse decays, and the triplets can account for the observed baryon asymmetry.    
We now turn to the quantitative study of leptogenesis for the case of the seesaw mechanism involving triplets. 

\subsection{The Boltzmann Equations and the Efficiency Factor}

There have been advances in several directions in the quantitative analysis of leptogenesis in type I seesaw models over the past few years, and more and more subtle effects are being taken into account such as spectator processes \cite{Buchmuller:2001sr}, thermal corrections \cite{Giudice:2003jh}, and flavor effects \cite{Barbieri:1999ma}, \cite{Nardi:2006fx}, \cite{Abada:2006ea}, \cite{Blanchet:2006be}, \cite{Blanchet:2007hv}. It would certainly be interesting to study effects such as thermal corrections to particle masses and CP violations or effects of flavor in the case of triplets, but we will limit ourselves to the ``traditional'' calculation for now. As long as the asymmetry is generated at temperatures above $10^{12}-10^{13}\,GeV$, which is where the Yukawa interactions between the heaviest leptons and the Higgs become important, the one flavor approximation should be a very good approximation. Even below this scale we get an idea whether triplets can generate the observed baryon asymmetry. While the effects of flavor tend to change the amount of asymmetry generated as a function of the light neutrino masses, these effects do not change the maximal asymmetry that can be generated~\cite{Blanchet:2006be}. We will further make the simplifying assumption that $\Delta L=2$ scattering processes can be ignored. As can be seen by simple dimensional analysis, this is expected to be a good approximation for triplet masses below some $10^{14}\,GeV$. Keeping these caveats in the back of our mind, we write down the relevant Boltzmann equations for our problem, working in the limit of hierarchical triplets. Elastic scattering processes occur at a rate faster than the rate of expansion of the universe at the time, meaning we can assume kinetic equilibrium. Furthermore, for the range of temperatures we are interested in, we can make the further, reasonably good approximation that the various particle species obey a Maxwell-Boltzmann distribution. The distribution functions of the various particles are then determined merely by the normalization, or the total number of particles. Since we work in a cosmological setting, it will be convenient for us to use comoving number densities. In other words we normalize the various number densities relative to the number density of photons at the time leptogenesis occurs appropriately redshifted. Under these assumptions the relevant dynamics is described by the following two coupled differential equations:
\begin{eqnarray}
\frac{dN_T}{dx}&=&-\frac{A(x)}{x^2}\left(N_T^2-N_{T,eq}^2\right)- D(x)\left(N_T-N_{T,eq}\right)\,,\\
\frac{dN_{B-L}}{dx}&=&-\varepsilon  D(x)\left(N_T-N_{T,eq}\right)-W(x)N_{B-L}\,.
\end{eqnarray}  
By $N_X$ we denote the ratio of $X$-particle density to the density of photons at the time of leptogenesis red-shifted by a factor $a^3$. $N_{X,eq}$ denotes the equilibrium value of this quantity. We have introduced the dimensionless variable $x=M/kT$. The CP violation in the decay of the lightest triplet is denoted by $\varepsilon$. The functions $A(x)$, $D(x)$, and $W(x)$ are given by:
\begin{eqnarray}
A(x)&=&\frac{2\alpha_2^2\zeta(3)}{\pi^2 M_1}\frac{M_1^2\langle\sigma v\rangle}{\alpha_2^2}\sqrt{\frac{45}{4\pi^3 G g_*}}\,,\\
D(x)&=&x\frac{\langle\Gamma\rangle}{ H_*}=\frac{\Gamma_0}{H_*}x \frac{K_1(x)}{K_2(x)}\,,\\
W(x)&=&\frac{1}{2}D(x)\frac{N_{T,eq}}{N_{\ell,eq}}=\frac{3}{4}D(x)x^2K_2(x)\,.
\end{eqnarray}  
The functions $K_1(x)$ and $K_2(x)$ are modified Bessel functions. The constant $\alpha_2$ is the usual $\alpha_2=\frac{g^2}{4\pi}$ with $g$ being the gauge coupling of $SU(2)_L$ evaluated at the triplet mass. $M_1$ denotes the mass of the lightest triplet. The Hubble constant at the time when the temperature equals the mass of the triplet, $H_*$, is given as
\begin{equation}
H_*=\sqrt{\frac{4\pi^3 G}{45}g_*}M_1^2\,.
\end{equation}
The quantity $g_*=106.75$ is the effective number of degrees of freedom of the standard model. In principle this number depends, of course, on temperature because the triplet is relativistic as long as the temperature is above its mass and becomes non-relativistic for temperatures below. Since leptogenesis occurs at roughly one tenth of the triplet's mass it is a good approximation to assume it is non-relativistic throughout the calculation and use a constant effective number of degrees of freedom. The cross section $\sigma$ is the cross section for triplets annihilating into gauge bosons and the average is a thermal average. To be explicit, it is given by
\begin{equation}
\langle\sigma v\rangle(x)=\frac{1}{M_1^2}\frac{x}{K_2^2(x)}\int\limits_1^\infty dy y^{3/2}\tilde\sigma(y)\beta(y)^2 K_1(2\sqrt{y}x)\,,
\end{equation}
where $K_1(x)$ and $K_2(x)$ are modified Bessel functions, $\beta(y)=\sqrt{1-1/y}$, and the dimensionless version of the cross section, $\tilde\sigma(y)$, is given by
\begin{equation}
\tilde\sigma(y)=4M_1^2\sigma(s)\,,
\end{equation} 
with $s=4M_1^2 y$ and the total annihilation cross section, $\sigma(s)$, for triplets at leading order given by the following expression:\footnote{Near threshold, one expects the cross section to be enhanced by non-perturbative effects such as Sommerfeld enhancement and logarithmic enhancement due to soft radiation. We checked numerically that these effects can be neglected in a first treatment but might have to be taken into account in a more detailed study especially for small $M_1$. The error introduced in the efficiency by ignoring these effects is at the per cent level for $K\gtrsim 1$ and increases to up to ten per cent for small $K$ for $M_1=10^{10}\,GeV$ and decreases with increasing $M_1$.}
\begin{multline}
\sigma(s)=\frac{\alpha_2^2\pi}{18s\beta}(3-\beta^2)+\frac{4\alpha_2^2\pi}{3s\beta}(3-\beta^2)\\
+\frac{2\alpha_2^2\pi}{9\beta^2s}\left(\ln\left(\frac{1+\beta}{1-\beta}\right)(21-6\beta^2-3\beta^4)-33\beta+17\beta^3\right)\,.
\end{multline}
The first term is the contribution of annihilation into Higgs, the second term the contribution of annihilation into quarks and leptons, and the last term is the contribution coming from annihilation into gauge bosons.
The rate $\Gamma_0$ is the decay rate of the triplet in its rest frame and is given by
\begin{equation}
\Gamma_0=\frac{(y^\dagger y)_{11}M_1}{32\pi}=\frac{1}{4\pi}\frac{M_1^2}{v^2}\sum\limits_i m_i|R_{1i}|^2\,,
\end{equation}
where $m_i$ are the masses of the light neutrinos and $R$ is the complex orthogonal matrix introduced in the previous section. The combination $\sum\limits_i m_i|R_{1i}|^2$ is often denoted as $\tilde{m}_1$.
In analogy to the singlet case, the solution can be expressed rather neatly in terms of the efficiency factor $\kappa_f$ as follows:
\begin{equation}
N_{B-L}(x)=-\frac{9}{4}\varepsilon \kappa(x)\,.
\end{equation}
The efficiency is normalized to be unity if all triplets decay in a CP violating way, and the quantity of interest is, of course $\kappa_f=\kappa(x\to\infty)$. The only difference to the singlet case at this point is the appearance of a factor of $\frac{9}{4}$ instead of the usual factor $\frac{3}{4}$,\footnote{This is just the ratio in number densities of a fermionic degree of freedom and a bosonic degree of freedom coming from the fact that we normalized to the number of photons} accounting for the fact that there are three times as many triplets in equilibrium as there are singlets. In Figure \ref{fig:efficiency}, we show the numerical results of our calculation of the efficiency factor as a function of the parameter $K=\Gamma_0/H_*=\frac{\tilde{m}_1}{10^{-3}\,eV}$. As comparison we show the efficiency for singlets in the same plot, but one should have in mind that the triplets produce three times the asymmetry singlets produce if the efficiencies and CP violation are the same.
\FIGURE[h]{
\includegraphics[width=5in]{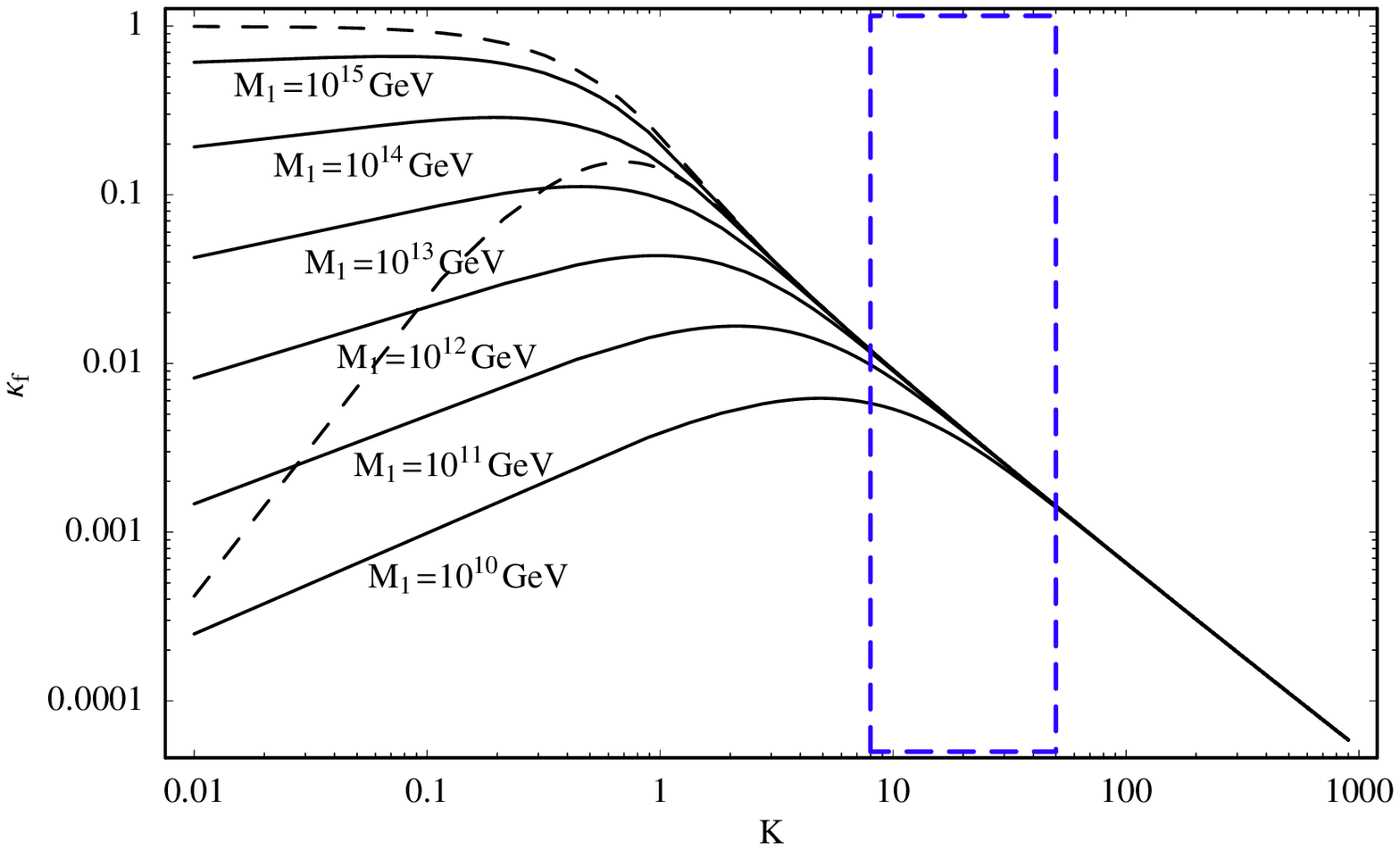}
\caption{The solid lines in this plot show the efficiency as a function of the parameter $K=\Gamma_0/H_*$ for triplets for a range of triplet masses $M_1$. The dashed lines represent the efficiency for singlets as a comparison. The upper and lower dashed lines correspond to singlets with an initial density equal to the equilibrium density and zero initial density, respectively.}\label{fig:efficiency}
}

\subsection{CP Violation}
Let us now turn to the calculation of the CP violation, $\varepsilon$, in the triplet decay. It is defined via
\begin{equation}
\varepsilon=\frac{\Gamma(T_1\to\ell \bar{H})-\Gamma(T_1\to\bar\ell H)}{\Gamma(T_1\to\ell \bar{H})+\Gamma(T_1\to\bar\ell H)}\,.
\end{equation}
The leading contribution comes from the interference of the tree and on-loop diagrams shown in Figure \ref{fig:CP}.
\FIGURE[h]{
\begin{tabular}{ccc}
\begin{fmffile}{one}    
  \fmfframe(1,7)(1,7){  
   \begin{fmfgraph*}(65,62) 
    \fmfleft{i1}        
    \fmfright{o1,o2}    
    \fmflabel{$T_1$}{i1}
    \fmflabel{$l$}{o2}
    \fmflabel{$\bar{H}$}{o1}
    \fmf{fermion}{i1,v1} 
    \fmf{fermion,tension=0.7}{o2,v1} 
    \fmf{dashes,tension=0.7}{v1,o1} 
   \end{fmfgraph*}
  }
\end{fmffile}&
\begin{fmffile}{two}    

\fmfcmd{
  style_def majorana expr p =
    cdraw p;
    cfill (harrow (reverse p, .5));
    cfill (harrow (p, .5))
  enddef;
}
\fmfcmd{
  style_def altmajorana expr p =
    cdraw p;
    cfill (tarrow (reverse p, .55));
    cfill (tarrow (p, .55))
enddef;}
  \fmfframe(1,7)(1,7){  
   \begin{fmfgraph*}(120,62) 
    \fmfleft{i1}        
    \fmfright{o1,o2}    
    \fmflabel{$T_1$}{i1}
    \fmflabel{$l$}{o2}
    \fmflabel{$\bar{H}$}{o1}
    \fmf{majorana}{i1,v1} 
    \fmf{fermion,tension=0.5,label=$l$}{v1,v3} 
    \fmf{dashes,tension=0.5,label=$H$}{v2,v1} 
    \fmf{fermion,tension=0.8}{o2,v2} 
    \fmf{dashes,tension=0.8}{v3,o1} 
    \fmf{altmajorana,tension=0.0,label=$T$}{v3,v2} 
   \end{fmfgraph*}
  }
\end{fmffile}&

\begin{fmffile}{three}  
  \fmfframe(1,7)(1,7){  
   \begin{fmfgraph*}(130,62) 
    \fmfleft{i1}        
    \fmfright{o1,o2}    
    \fmflabel{$T_1$}{i1}
    \fmflabel{$l$}{o2}
    \fmflabel{$\bar{H}$}{o1}
    \fmf{fermion}{i1,v2} 
    \fmf{fermion,left=0.5,tension=0.3,label=$l$}{v3,v2} 
    \fmf{dashes,left=0.5,tension=0.3,label=$H$}{v2,v3} 
    \fmf{fermion}{v3,v1} 
    \fmf{fermion,tension=0.5}{o2,v1} 
    \fmf{dashes,tension=0.5}{v1,o1} 
   \end{fmfgraph*}
  }
\end{fmffile}
\end{tabular}
\caption{This plot shows the tree and one-loop diagrams whose interference gives rise to the leading contribution to the CP violation in the triplet decay.}\label{fig:CP}
}\\

It is given by
\begin{multline}
\varepsilon=-\frac{\text{Im}\left(y_{i1}y^*_{mn}y^*_{in}y_{m1}\right)}{32\pi\sum_j|y_{j1}|^2}\frac{M_n}{M_1}\left(1+\left(1+\frac{M_n^2}{M_1^2}\right)\ln\left(\frac{M_n^2}{M_1^2+M_n^2}\right)\right)\\+\frac{\text{Im}\left(y_{i1}y^*_{mn}y_{in}y^*_{m1}\right)}{32\pi\sum_j|y_{j1}|^2}\frac{M_1^2}{M_1^2-M_n^2}+\frac{\text{Im}\left(y_{i1}y^*_{mn}y^*_{in}y_{m1}\right)}{32\pi\sum_j|y_{j1}|^2}\frac{M_1M_n}{M_1^2-M_n^2}\,,
\end{multline}
where a summation over $i,m$ and $n$ is implied, and $M_n$ is the mass of the triplet in the loop.\footnote{We note that compared to the singlet case the first term has opposite sign. This sign can be traced to the minus sign in the identity ${{\tau^A}^a}_b{{\tau^A}^c}_d=(\delta^a_d\delta_b^c-1/2\delta_b^a\delta_d^c)/2$. As a side remark, this also implies that the vertex contribution to the CP violation is $1/N$ suppressed in a large $N$ counting, but for $N=2$ this is of course not terribly relevant.} 
In the limit of hierarchical triplets, and expressed in terms of observable quantities rather than the Yukawa couplings, this becomes
\begin{equation}
\varepsilon=-\frac{1}{8\pi}\frac{M_1}{v^2}\frac{\sum\limits_im_i^2\text{Im}\left(R_{1i}^2\right)}{\sum\limits_i m_i|R_{1i}|^2}\,.
\end{equation}
We note that once expressed in terms of physical quantities such as neutrino masses and the entries of the complex orthogonal matrix $R$, this is smaller than the CP violation in the case of singlets by a factor three.\\ 
As first shown in \cite{Davidson:2002qv}, the orthogonality of the matrix $R$ can be used to derive an upper bound on the magnitude of the CP violation given by
\begin{equation}
|\varepsilon|\leq\frac{1}{8\pi}\frac{M_1(m_3-m_1)}{v^2}\,,
\end{equation}
where $m_3$ and $m_1$ are the masses of the heaviest and lightest of the light neutrinos, respectively. Thinking of $m_3-m_1$ as $(m_3^2-m_1^2)/(m_3+m_1)$, it is apparent that this becomes maximal for hierarchical neutrinos with $m_1=0$ and $m_3=\sqrt{\Delta m_\text{atm}^2}\approx 5\times10^{-2}\,eV$. The maximal amount of CP violation is then given by
\begin{equation}
\varepsilon^\text{max}=3.3\times 10^{-7}\left(\frac{M_1}{10^{10}\,GeV}\right)\left(\frac{\sqrt{\Delta m_\text{atm}^2}}{0.05\,eV}\right)\,.
\end{equation}
\subsection{The Baryon Asymmetry}
All that is left now is to combine the result from the last two subsections to obtain the final baryon asymmetry. It is given by
\begin{equation}
\eta=\frac94 d \kappa_f \epsilon\,,
\end{equation}
where $d=\frac{28}{79}\frac{86}{2387}$. The first factor converts from the $B-L$ asymmetry we have calculated to a baryon asymmetry. The second factor takes into account that the photon temperature, and hence the number of photons, increases every time a particle species goes out of equilibrium as the temperature drops below its mass, which dilutes the baryon-to-photon ratio. There could be additional entropy production between the time leptogenesis occurs and today which would dilute the baryon asymmetry further. All we can therefore be certain of is that the asymmetry generated must be greater than or equal to the asymmetry that is observed today
\begin{equation}
\eta=\frac94 d \kappa_f(M_1,K) \epsilon(M_1)\geq6.2\times 10^{-10}\,.
\end{equation} 
We present it as a lower bound on the mass of the lightest of the triplets. This is shown in Figure \ref{fig:massbound}.
\FIGURE[h]{
\includegraphics[width=5in]{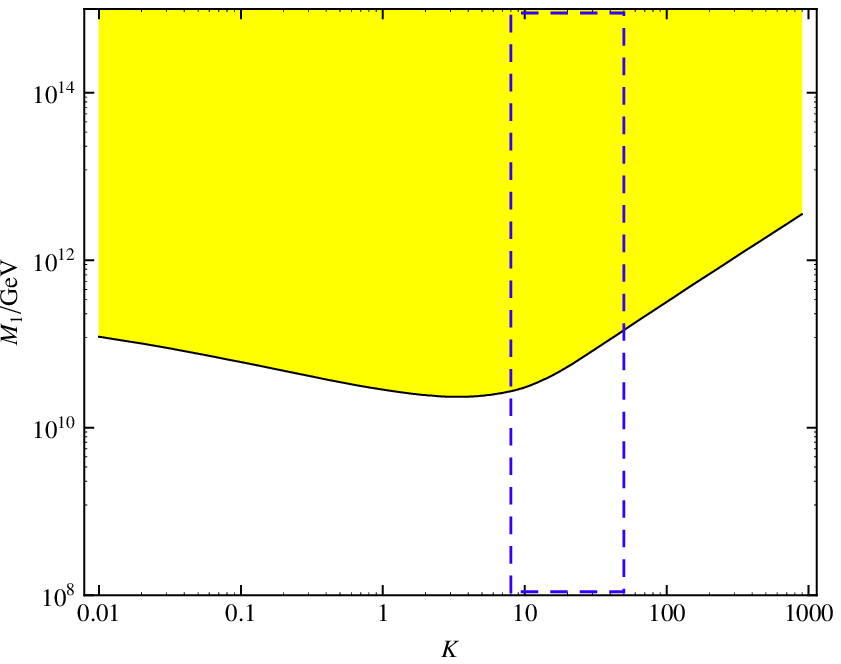}
\caption{This solid line in this plot shows some the efficiency as a function of the parameter $K=\Gamma_0/H_*$ for triplets. The dashed lines represent the efficiency for singlets as a comparison. The upper and lower dashed lines correspond to singlets with an initial density equal to the equilibrium density and zero initial density, respectively.}\label{fig:massbound}
}
For the range of neutrino masses that seems to be realized in nature, the bound to a good approximation takes the form
\begin{equation}
M_1>(1.5\pm0.1)\times 10^{11}\,GeV\left(\frac{K}{50}\right)^{1.1}\gtrsim 9\times 10^{10}\,GeV \left(\frac{K}{50}\right)^{1.1}\,,
\end{equation}
where the second inequality is the commonly quoted 3$\sigma$ bound. The quoted error is a combination of statistical uncertainties in the determination of the baryon asymmetry from the CMB taken from \cite{Komatsu:2008hk} and uncertainties in the measurements of $\Delta m_\text{atm}^2$ taken from \cite{Ashie:2004mr}, which is the error commonly quoted in this context. The systematic uncertainties in the determination of the baryon asymmetry are much larger than the statistical ones, and there are various theoretical uncertainties that have not been included into the estimate. We should therefore take it with a grain of salt. 

From Figure \ref{fig:massbound} we see that there is an absolute lower limit on the mass of the lightest triplet if we require successful generation of the baryon asymmetry in the case of hierarchical triplets.
\begin{equation}
M_1>(2.3\pm0.2)\times 10^{10}\,GeV\gtrsim1.7\times 10^{10}\,GeV\,,
\end{equation}
where the second inequality again indicates the 3$\sigma$ bound. This is somewhat stronger than the bound found in the singlet case.\footnote{Just as in the singlet case the bound can be lowered substantially if the triplets are no longer hierarchical, one takes the effects of flavor into account, or if one resorts to resonant leptogenesis \cite{Pilaftsis:2003gt}.}
One should of course keep in mind that not only does the lightest triplet have to be heavier than this bound for leptogenesis to work, but the universe must also reheat to temperatures higher than this after inflation for successful leptogenesis to occur in our model. \\
In this context, it is worth pointing out that, while in theories with a low scale of supersymmetry breaking the high reheat temperature required for successful leptogenesis can be dangerous because of the gravitino problem~\cite{Weinberg:1982zq},~\cite{Linde:1984pg},~\cite{Falomkin:1984eu}, there is no gravitino problem in our case, because supersymmetry is assumed to be broken at a very high scale.

\section{Realization in the Context of GUT Theories}\label{unification}
If there is a singlet state present in addition to the standard model degrees of freedom, the matter content fits rather nicely into the ${\bf 16}$ of $SO(10)$. In this case, in the absence of low energy supersymmetry, the gauge couplings do not unify and one is lead to study GUT theories with a low supersymmetry breaking scale. \\
If a triplet state is present, the running of $\alpha_2$ is modified, and unification can be achieved in the absence of low energy supersymmetry provided the mass spectrum of triplets and their unification partners is appropriate.\\ 
We will show how this can be arranged in a technically natural way in the example of the group $SU(5)$. It could easily be extended to the other popular GUT groups.  
\subsection{The Model and the Actors II}
To achieve grand unification, we have to include grand unification partners for the triplets. The triplets are components of Weyl fermions, $\Sigma_i$, transforming in the adjoint representation of $SU(5)$\footnote{Grand unified theories with gauge group $SU(5)$ extended by a single adjoint fermion have been studied in \cite{Bajc:2006ia}, \cite{Bajc:2007zf},\cite{Dorsner:2006fx} and theories with gauge group $SU(5)$ extended by a single adjoint and a Higgs in the ${\bf 45}$ in \cite{Perez:2007rm}.} and the Lagrangian of the effective theory describing the system just above the scale at which unification occurs takes the form:\footnote{We do not address the doublet triplet splitting. There will be additional fields needed to address this issue that we suppress for now. }
\begin{equation}
\mathcal{L}=\mathcal{L}_{\text{kinetic}}+\mathcal{L}_{\text{Yukawa}}-V(H_5,H_{24},H_{75} )
\end{equation}
with:
\begin{eqnarray}\label{Lagr}
\mathcal{L}_{\text{kinetic}}&=& -\frac{1}{2g_5^2}\text{Tr}\left(F_{\mu\nu}F^{\mu\nu}\right)-\chi_{i\,A}^\alpha iD_{\alpha\dot{\alpha}}\overline{\chi}^{\dot\alpha\,A}_i-\psi_{i}^{\alpha\,AB} iD_{\alpha\dot{\alpha}}\overline{\psi}^{\dot\alpha}_{i\,AB}\\\nonumber
&&+\left(D_\mu H_5\right)^\dagger_A\left(D^\mu H_5\right)^A+{{\left(D_\mu H_{24}\right)^\dagger}^{A}}_{B}{{\left(D^\mu H_{24}\right)}^{B}}_{A}\\\nonumber
&&+{{\Sigma_{i}^{\alpha}}^{A}}_{B} iD_{\alpha\dot{\alpha}}{{\overline{\Sigma}^{\dot\alpha\,B}_{i}}}_{A}+                                                                \frac{1}{2}M_i\left({{\Sigma_{i}^{\alpha}}^{A}}_{B}{{\Sigma_{\alpha\,i}}^{B}}_{A}+h.c.\right)\\\nonumber
&&+{\left(D_\mu H_{75}\right)^\dagger}^{AB}_{CD}{\left(D^\mu H_{75}\right)}^{CD}_{AB}
\end{eqnarray}
and:
\begin{eqnarray}
\mathcal{L}_{\text{Yukawa}}&=&-g_{ij}\psi_i^{AB}\chi_{j\,B}{H_5}^\dagger_A-\tilde{g}_{ij}\epsilon_{ABCDE}\psi_i^{AB}\psi_j^{CD} H_5^E\\\nonumber
&&-G_{ij}{{\Sigma_{i}^{\alpha}}^{A}}_{B}{{\Sigma_{\alpha\,j}}^{B}}_{C}{{H_{24}}^{C}}_{A}-\tilde{G}_{ij}{{\Sigma_{i}^{\alpha}}^{A}}_{C}{{\Sigma_{\alpha\,                j}}^{B}}_{D}{H_{75}}^{CD}_{AB}\\\nonumber
&&-y_{ij}\chi^\alpha_{i\,A}{{\Sigma_{\alpha\,j}}^{A}}_{B}H_5^B+h.c.
\end{eqnarray}
where the covariant derivative is defined as usual:
\begin{equation}
D_\mu\mathcal{O}^{AB\dots}_{CD\dots}=\partial_\mu\mathcal{O}^{AB\dots}_{CD\dots}-i{A_\mu^A}_E\mathcal{O}^{EB\dots}_{CD\dots}-                                          i{A_\mu^B}_E\mathcal{O}^{AE\dots}_{CD\dots}+i{A_\mu^E}_C\mathcal{O}^{AB\dots}_{ED\dots}+i{A_\mu^E}_D\mathcal{O}^{AB\dots}_{CE\dots}+\dots
\end{equation}
The field $\chi$ denotes the usual left-handed $\overline{{\bf 5}}$, $\psi$ is the left-handed ${\bf 10}$, $H_5$ is the $5$-Higgs containing the SM Higgs, $H_{24}$ is the usual Higgs in the adjoint of $SU(5)$ whose vacuum expectation value breaks $SU(5)$ down to the standard model gauge group in ordinary $SU(5)$ grand unified theories. 

The new fields are the Higgs in the ${\bf 75}$, $H_{75}$,\footnote{The reason for including a Higgs in the ${\bf 75}$ will become clear in the next section when we talk about constraints from grand unification. In brief it is required if we want a renormalizable model with the mass spectrum for triplets and octets needed for unification.}
 and the fermion $\Sigma$ that transforms in the adjoint and contains the triplet. To be completely explicit, let us specify how $\Sigma$ is related to the triplet:
\begin{equation}
\Sigma=\sqrt{2}\left(\begin{array}{cc}
O-\frac{2}{\sqrt{60}}N\mathbf{1}_{3\times 3} & M_1\\
M_2 & T+\frac{3}{\sqrt{60}}N\mathbf{1}_{2\times 2}
\end{array}\right)
\end{equation}
Under $SU(3)\times SU(2) \times U(1)$ $O$ transforms as $({\bf 8},{\bf 1})_0$ and will from now on be referred to as the octet, the triplet, $T$, transforms according to $({\bf 1},{\bf 3})_0$, $M_1$ as $({\bf 3},{\bf 2})_{-\frac{5}{6}}$, and $M_2$ as              $(\overline{{\bf 3}},{\bf 2})_{\frac{5}{6}}$.
The fields $M_1$ and $M_2$ are typically required to have masses close to the GUT scale. Different from what one might expect this is not required to evade the experimental limits on proton decay. Since the fields are fermions and the leading baryon number violating operators generated in the low energy theory by integrating them out are dimension seven. So the bounds on the proton lifetime lead to a much weaker lower bound \cite{Weinberg:1980bf} on the mass of around $10^{10}\,GeV$. Instead these particles have to be heavy for the following reason. The octet is long-lived.\footnote{The decay involves a virtual Higgs triplet whose mass is of order $M_\text{GUT}$. In addition, the Yukawa couplings involved in its decay are related to the Yukawa couplings governing the triplet decay due to unification and are forced to be small by the mass scale of the light neutrinos as we shall see in Section \ref{numass}.} In order for it to decay before nucleosynthesis it must have a mass of a at least around $2.5\times 10^{10}\,GeV$. Together with unification this puts a lower bound on the mass of the fields $M_1$ and $M_2$ of about $10^{14}\,GeV$.

\subsection{Constraints from Unification of Coupling Constants}

Before we write down the conditions that are forced on us by our bias that the gauge couplings should unify, a few remarks about the logic seem appropriate.

In the usual discussion of the unification of gauge coupling constants one takes two of the coupling constants of the standard model at the weak scale as input parameters and predicts the value of the grand unification scale, $M_\text{GUT}$, as well as the coupling constant $\alpha_\text{GUT}$ of the grand unified theory at this scale. Assuming unification one can then predict the third coupling of the standard model. For a long time the two low energy couplings used were $\alpha_S(M_Z)$ and $\alpha_{EM}(M_Z)$ and the quantity predicted was $\sin^2\theta_W$. Since $\alpha_S(M_Z)$ is the coupling with the greatest uncertainty it has become more popular to use $\alpha_{EM}$ and $\sin^2\theta_W$ to predict $\alpha_S(M_Z)$ but the idea remains the same.

In our case the logic is somewhat different because the model has additional parameters and is less restricted. Instead of predicting the energy scale of grand unification, which is impossible due to the intermediate mass scales we are introducing into the theory, we get to choose the GUT scale. We take it to be $M_\text{GUT}=2\times 10^{16}\,GeV$ to ensure a long enough lifetime for the proton. We then use $\alpha_1(M_Z)$ to predict the gauge coupling of the grand unified theory at the unification scale. This is straightforward since the only new, potentially light, fields carrying hypercharge are the fields $M_1$ and $M_2$. Since their masses are close to $M_\text{GUT}$, the running of $\alpha_1(\mu)$ is essentially identical with its running in the standard model and we simply have $\alpha_\text{GUT}(M_\text{GUT})=\alpha_1(M_\text{GUT})$. 
Given $\alpha_\text{GUT}$ and $M_\text{GUT}$, assuming grand unification, and imposing that the values of the other two gauge couplings of the standard model at the weak scale be compatible with their experimental values we find that the masses of the triplets and octets have to satisfy the following relations:\footnote{These relations are valid assuming the fields $M_1$ and $M_2$ have masses of order $M_\text{GUT}$ and do not contribute to the running of the couplings. If their masses are somewhat below the GUT scale the effect is that the masses of both the octets and the triplets decrease. For details see Appendix \ref{GUT}}
\begin{equation}\label{tmass}
M_{T_1}M_{T_2}M_{T_3}=8.54\times 10^{30}\,GeV^3\left(\frac{2\times 10^{16}\,    GeV}{M_\text{GUT}}\right)^{2.45}\,,
\end{equation}
and similarly for the octets:
\begin{equation}\label{omass}
M_{O_1}M_{O_2}M_{O_3}=1.60\times 10^{38}\,GeV^3\left(\frac{2\times 10^{16}\,    GeV}{M_\text{GUT}}\right)^{2.55}\,.
\end{equation}

If we assume that the Majorana masses $M_i$ in eq. \eqref{Lagr} are of the same order of magnitude as the grand unified scale, $M_\text{GUT}$, we would expect both the triplet as well as the octet to have masses of that same order of magnitude. However, once both the Higgs in the ${\bf 24}$ and the Higgs in the ${\bf 75}$ acquire a vacuum expectation value and the grand unified symmetry is broken, we get an additional contribution to their masses from their Yukawa interactions and we assume that the couplings are chosen such that the above relations are satisfied. We would like to emphasize that while this requires a particular choice of values for the parameters in our model, this choice is stable under quantum corrections. 

For completeness let us mention that the singlet, $N$, as well as the fields $M_1$ and $M_2$ will then have a mass of order $M_i\propto M_\text{GUT}$.  

For more details about the derivation of the conditions \eqref{tmass}, \eqref{omass}, we refer the interested reader to Appendix \ref{GUT}.

One might worry whether the theory above the grand unified scale is still asymptotically free, which is a nice feature of both the non-SUSY and the SUSY SU(5) grand unified theories. It turns out that this is in fact no longer the case in this model. However, even though the coupling $\alpha_\text{GUT}$ will increase as the energy scale increases beyond $M_\text{GUT}$, the theory remains perturbative all the way up to the Planck scale where we expect new physics to become relevant.

\section{Conclusions and Outlook}
We presented a model that in addition to the standard model degrees of freedom contains $SU(2)_L$-triplet fermions as well as $SU(3)_c$-octet fermions at intermediate energies and further degrees of freedom at the GUT scale. \\

The model provides a neutrino mass matrix consistent with current experiments and is capable of generating the observed baryon asymmetry of the universe provided that the lightest triplet has a mass around or above $10^{10}\,GeV$. For triplet masses of that same order of magnitude, the model can lead to unification of gauge coupling constants in the absence of low energy supersymmetry without dangerous proton decay provided that the product of the octets' masses is around $10^{38}\,GeV^3$, and the remaining unification partners of the triplets have masses around the GUT scale.  \\
In the context of $SU(5)$, we showed that this spectrum can be obtained in a technically natural way.\\

We showed, that there are hardly any constraints on the triplets' masses from experimental bounds on flavor changing neutral currents in the lepton sector, from bounds on the electron electric dipole moment, or from precision electroweak measurements. \\
It is therefore conceivable that one of the triplets is light enough to be seen at the LHC \cite{Franceschini:2008pz}, or more easily at the ILC. Since this is of course not required to be the case by anything we know of, generically the model is, however, a realization of the nightmare scenario that the LHC will discover a Higgs and nothing else. This can of course be ``accomplished'' in easier ways, but one should keep in mind the virtues of this model over other scenarios of that type such as providing neutrino masses, successful leptogenesis, and grand unification.\\     

As it stands our model does not provide a natural dark matter candidate, but it could easily be modified to accommodate axionic dark matter.\\
Another way in which it can be modified is by imposing a symmetry under which the triplet and its unification partners are charged while the standard model particles are neutral. In this case the lightest triplet becomes a natural candidate for dark matter, which could possibly be seen at the LHC. \\
To generate neutrino masses, additional states charged under this new symmetry with the quantum numbers of the standard model Higgs and their unification partners have to be introduced.\\ 
Under certain circumstances the baryon asymmetry of the universe can then be produced by late decaying octets, while still maintaining grand unification. We leave this for future work.\\

\section{Acknowledgements}
We would like to thank Sonia Paban for helpful conversations. R.F. would also like to thank Yanou Cui, Matt Reece, and Korneel van den Broek for various entertaining discussions.
This research was supported in part by the National Science Foundation under Grant No. PHY-0455649. The work of R.F. is also funded in part by the Kavli Institute for Theoretical Physics under the National Science Foundation Grant No. PHY05-51164.

\appendix

\section{Appendix: Grand Unification -- A More Detailed Discussion}\label{GUT}
Let us look at the evolution of the couplings in somewhat more detail. As mentioned in Section \ref{unification} we get to pick $M_\text{GUT}$ and we choose it to be $2\times 10^{16}\,GeV$ to evade bounds on the proton lifetime. We will then evolve $\alpha_1(M_Z)$ up to the grand unification scale and hence predict $\alpha_\text{GUT}$. Requiring that the values of $\alpha_2(M_Z)$ and $\alpha_S(M_Z)$ obtained from the renormalization group be compatible with the experimentally observed ones will put bounds on the masses of triplets and octets. We will work in the $\overline{\text{MS}}$ scheme. In this scheme the current experimental values for $\alpha_{EM}(M_Z)$, $\sin^2\theta_W(M_Z)$, and $\alpha_S(M_Z)$ taken from \cite{Yao:2006px} give rise to the following values for $\alpha_1(M_Z)$, $\alpha_2(M_Z)$, and $\alpha_S(M_Z)$:
\begin{eqnarray}
\alpha_1^{-1}(M_Z)&=&59.00\pm0.02\\\nonumber
\alpha_2^{-1}(M_Z)&=&29.59\pm0.02\\\nonumber
\alpha_S^{-1}(M_Z)&=&8.50\pm0.14
\end{eqnarray}
At one loop the renormalization group equations predict a grand unified coupling of:
\begin{equation}
\alpha_\text{GUT}^{-1}(M_\text{GUT})=\alpha_1^{-1}(M_Z)+\frac{1}{2\pi}b_1\ln\frac{M_\text{GUT}}{M_Z}=37.45\,,
\end{equation}
where $b_1=-41/10$ is the standard model value.
Using this as initial condition in the one-loop renormalization group equations for the two remaining couplings we find the following condition on the masses:
\begin{eqnarray}
\alpha_2^{-1}(M_Z)&=&\alpha_1^{-1}(M_Z)-\frac{1}{2\pi}(b_2-b_1)\ln\frac{M_\text{GUT}}{M_Z}+\frac{2}{3\pi}\ln\left(\frac{M_\text{GUT}^3}{M_{T_1}M_{T_2}M_{T_3}}\right)\\
\alpha_S^{-1}(M_Z)&=&\alpha_1^{-1}(M_Z)-\frac{1}{2\pi}(b_3-b_1)\ln\frac{M_\text{GUT}}{M_Z}+\frac{1}{\pi}\ln\left(\frac{M_\text{GUT}^3}{M_{O_1}M_{O_2}M_{O_3}}\right)
\end{eqnarray}
where again $b_2=19/6$ and $b_3=7$ are the standard model values.
This leads to the conditions on the masses of triplets and octets quoted in Section \ref{unification}:
\begin{eqnarray}
M_{T_1}M_{T_2}M_{T_3}&=&8.54\times 10^{30}\,GeV^3\left(\frac{2\times 10^{16}\,    GeV}{M_\text{GUT}}\right)^{2.45}\,\\
M_{O_1}M_{O_2}M_{O_3}&=&1.60\times 10^{38}\,GeV^3\left(\frac{2\times 10^{16}\,    GeV}{M_\text{GUT}}\right)^{2.55}\,.
\end{eqnarray}
If we lower the mass of $M_1$ and $M_2$ significantly below the GUT scale we should take it into account. This gives rise to the following modified conditions:
\begin{eqnarray}
\frac{M_{T_1}M_{T_2}M_{T_3}}{M_{M_1}M_{M_2}M_{M_3}}&=&1.07\times 10^{-18}\,\left(\frac{2\times 10^{16}\,    GeV}{M_\text{GUT}}\right)^{5.45}\,\\
\frac{M_{O_1}M_{O_2}M_{O_3}}{M_{M_1}M_{M_2}M_{M_3}}&=&2.00\times 10^{-11}\,\left(\frac{2\times 10^{16}\,    GeV}{M_\text{GUT}}\right)^{5.55}\,.
\end{eqnarray}
We see that this condition is invariant under a rescaling of the masses of all particles by a common factor. This is something one might have guessed since a common rescaling of the masses of all the components in a representations of $SU(5)$ will not change whether or at what scale unification occurs. 
It is hence quite trivial to take into account if the masses of the fields $M_1$ and $M_2$ are slightly below the GUT scale. A convenient way to make this apparent is to rewrite the condition as:
\begin{eqnarray}
M_{T_1}M_{T_2}M_{T_3}\left(\frac{M_\text{GUT}^3}{M_{M_1}M_{M_2}M_{M_3}}\right)&=&8.54\times 10^{30}\,GeV^3\left(\frac{2\times 10^{16}\,    GeV}{M_\text{GUT}}\right)^{2.45}\,\\
M_{O_1}M_{O_2}M_{O_3}\left(\frac{M_\text{GUT}^3}{M_{M_1}M_{M_2}M_{M_3}}\right)&=&1.60\times 10^{38}\,GeV^3\left(\frac{2\times 10^{16}\,    GeV}{M_\text{GUT}}\right)^{2.55}\,.
\end{eqnarray}

As far as the strength of the coupling of the grand unified theory is concerned, it should be clear that the smaller the masses the larger the coupling of the grand unified theory and we might be worried that our theory becomes strongly coupled before the Planck scale if the new particles are too light, especially since the theory is not asymptotically free. However, as mentioned in Section \ref{unification}, there is a fairly stringent lower bound on the mass of the octets coming from nucleosynthesis. This bound translates to a lower bound for the $M_1$ and $M_2$ fields of around $10^{14}\,GeV$, incidentally ensuring that the theory remains weakly coupled all the way to the Planck scale. 

\newpage
\bibliographystyle{JHEP}  
\bibliography{triplets}   

\end{document}